# Electronic Structure Calculation of AsSiTeB/SiAsBTe nanostructures using the Density Functional Theory.

Ankit Kargeti[1, a)], Ravikant Shrivastav[1] and Dr.Tabish Rasheed[1]

[1]*Department of Applied Sciences, School of Engineering and Technology , BML Munjal University Gurgaon, Haryana 122413 India*

a) *ankitkageti@gmail.com*

**Abstract:** The electronic structure calculation for the nanoclusters of (AsSiTeB/SiAsBTe) quaternary semiconductor alloy belonging to the (III-V Group elements) is performed. The two clusters one in the linear form and the other in the bent form have been studied under the framework of Density Functional Theory (DFT) using the B3LYP functional and LANL2DZ basis set with the software packaged GAUSSIAN 16 . We have discussed the Optimised Energy, Frontier Orbital Energy Gap in terms of HOMO-LUMO, Dipole Moment, Ionisation Potential, Electron Affinity, Binding Energy and Embedding Energy value in the research work and we have also calculated the Density of States (DoS) spectrum for the above quaternary system for two nanoclusters. The application of these compounds or alloys are mainly in the Light emitting diodes.

Motivation for this research work is to look for electronic and geometric data of nanocluster (AsSiTeB/SiAsBTe).

**Keywords:** Density functional theory (DFT), III-V Group Elements, HOMO-LUMO, Ionisation Potential, Electron Affinity.

**Introduction:**

The III-V Group elements of the periodic table belong to the semiconducting region in which Arsenic Silicon heterostructure alloys provide the opportunities to form variety of opto-electronic devices. The diode lasers, photo-detectors, high efficiency solar cells, light emitting diodes are of the potential application for these chosen system whose synthesis can provide the substitution for present Gallium Nitride based system [1-3]. The heterostructures are used to form low dimensional electronic systems like quantum wells or quantum dots by confining the electrons and holes. Experimentally one of the method for these heterostructures preparation is of Metalorganic vapor phase epitaxy (MOVPE)[4-5].

In our work we have optimized the nanoclusters (AsSiTeB/SiAsBTe) using B3LYP Functional with LANL2DZ basis set under the Density Functional Theory. Density Functional Theory is a widely used method to investigate the structural and electronic properties of nanoclusters[6-9]. In our calculation we have used the B3LYP as the hybrid functional which stands for "Becke, 3-parameter, Lee–Yang–Parr". The Becke is the exchange functional and the correlation part is of Lee,Yang,Parr.

$$E_{XC}^{B3LYP} = E_X^{LDA} + \alpha_0 (E_X^{HF} - E_X^{LDA}) + \alpha_X (E_X^{GGA} - E_X^{LDA}) + E_C^{LDA} + \alpha_C (E_C^{GGA} - E_C^{LDA}),$$

where, $\alpha_0 = 0.20, \alpha_X = 0.72 \text{ and } \alpha_C = 0.81, E_X^{GGA}$ and $E_C^{GGA}$

are generalized gradient approximations. The Becke 88 exchange functional and the correlation functional of Lee, Yang and Parr for B3LYP, and $E_C^{LDA}$ is the VWN local-density approximation to the correlation functional.

**Computational Details:**

The two nanoclusters one in the Linear form and the other in the Bent form are optimized under the framework of the Density Functional Theory using B3LYP functional with LANL2DZ basis set which has been taken in view of the other theoretical studies performed for different alloy systems [11-14]. Since the two elements of AsSiTeB/SiAsBTe (arsenic, silicon ) are present in the cluster therefore LANL2DZ basis set is the suitable one and is applicable to the elements H, Li-La and Hf-Bi therefore that would be useful for optimizing the clusters. The optimized energy calculated for both the structures in which Bent structure is more stable E(-42.86 Hartree) as compared to the linear structure because its energy is E(-42.73 Hartree) on the potential energy surface. Therefore it shows that geometrical orientation is one of the important factor for the stability of the cluster. We have calculated the Frontier Orbital Energy gap value by the difference of the HOMO- LUMO value. The Band gap energy calculation for the bent structure is (2.77 eV) and for linear structure is (2.37 eV). Therefore the linear structure is having more preference over the bent structure for the process of electron transfer. Dipole moment for the bent structure is ( 0.71 Debye) and linear structure is ( 0.97 Debye) which shows the linear structure is more preferable for electron dissociation. Then Electron Affinity for bent structure is ( 3.35 eV) and for linear structure is ( 4.25eV) which shows that the electron acceptance power of the linear structure is more than the bent structure. The Ionisation potential for the bent structure is ( 6.16 eV) and for the linear structure is ( 6.63 eV) which shows that the electron ejection from the outermost molecular orbital of the bent structure is more easy than the linear structure. Binding energy for bent structure is ( 14.05 eV) and for linear structure is ( 10.53 eV ) which shows the bent structure is more tightly bonded than the linear structure.

**Results and discussion:**

**Optimised geometry-**

In this work we have presented the two different nanoclusters of the (AsSiTeB/SiAsBTe) semiconducting alloys one is the linear and the other one is bent structure. The bent structure is the two dimensional structure where we have arranged the linear structure in such a manner by which it could form a bent geometry. Figure 1 shows the optimized geometry of the linear and bent structure.

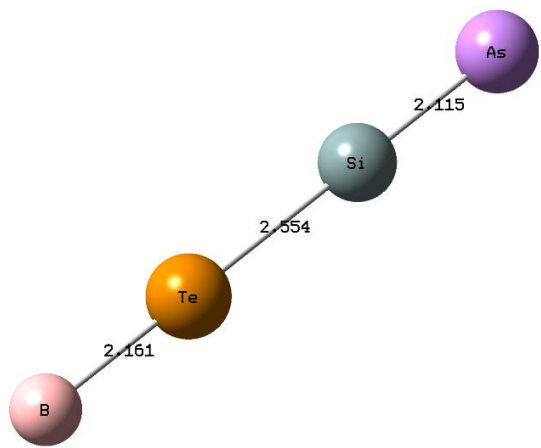 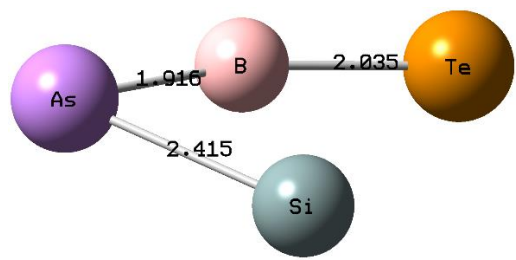

**FIGURE 1A. Optimised Geometry of Linear (AsSiTeB) Nanocluster**

**FIGURE 1B Optimised Geometry of Bent(SiAsBTe) Nanocluster**

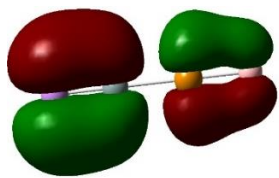 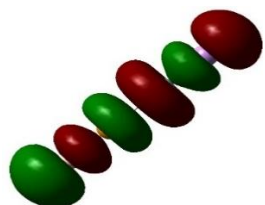 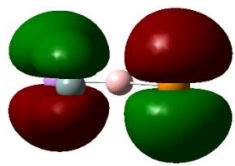 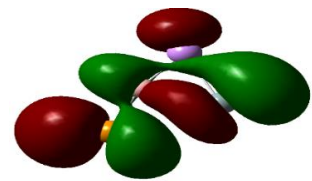

**FIGURE 2A. Linear (AsSiTeB) HOMO**

**FIGURE 2B. Linear (AsSiTeB) LUMO**

**FIGURE 2C. Bent (SiAsBTe) HOMO**

**FIGURE 2D. Bent (SiAsBTe) LUMO**

The optimized geometry of linear nano-cluster is shown in figure 1A where the optimized bond length between the Arsenic atom and Silicon atom is the 2.11 Å, the bond length for Silicon atom to the Tellurium atom is the 2.55Å and the bond length of the Tellurium to Boron atom is 2.16Å.

For the figure 1B the optimized geometry is of bent structure of the nano-cluster of (SiAsBTe) where the equilibrium bond length between Silicon atom to Arsenic atom is 2.41Å, for the Arsenic to the Boron atom is 1.91 Å, and for the Boron atom to the Tellurium atom is 2.03Å.

**Electronic Properties -** The band gap energy of the frontier orbital is given by the 2.77 eV for the bent structure and 2.37 eV for the linear structure which is imaged in figure 2 above. This shows that the cluster with changes in geometrical orientation effects the process of charge transfer of the compound.

**Chemical Hardness -** From the principal of maximum hardness which states that if the material is having low value of Energy band gap then this material falls under the category of soft matter which means it shows reactive nature towards chemical reaction while for the higher value of Energy band gap the material is non reactive towards chemical reaction hence called hard material. For the linear structure this vaule of hardness is (1.18 eV) and for the bent structure this value is (1.38 eV). Therefore the bent structure is more hard material as compared to the linear structure.

**Electronegativity** – Electronegativity of the material of the linear structure is around 5.46 eV and for the bent structure it is 4.79 eV which shows that Linear structure is more efficient in accepting the electron from the other donating species than the bent structure.

**Electrophilicity-** It defines the electron acceptor power of the molecule. Therefore the linear structure with high value of electrophilicity (12.51eV) than the bent structure which is (8.21eV) makes linear structure to be the highly electronegative which is also confirmed from our electronegativity equation.

**Embedded Energy** – It defines the embedding of any foreign atom in the molecule. For both the structures this energy is quite high as compared to the optimized energy of the clusters. Therefore its hard to embed foreign atom in the cluster.

**TABLE 1.** Optimized energy (O.E.), HOMO- LUMO Energy levels, Energy band gap (Eg), Electron affinity (EA), Ionization potential (IP), Binding energy (BE), Embedded Energy (EE), Dipole moment (ρ), values of the studied compounds obtained at B3LYP/LANL2DZ level.

| Cluster | O.E. (Hartree) | $E_{LUMO}$ (eV) | $E_{HOMO}$ (eV) | Eg (eV) | EA (eV) | IP (eV) | BE (eV) | EE (eV) | ρ (Debye) |
|---|---|---|---|---|---|---|---|---|---|
| Linear (AsSiTeB) | -42.7344 | -4.2809 | -6.6565 | 2.3755 | 4.2599 | 6.6322 | 10.5322 | 2.6331 | 0.9710 |
| Bent (SiAsBTe) | -42.8643 | -3.4085 | -6.1868 | 2.7783 | 3.3875 | 6.1625 | 14.0538 | 3.5135 | 0.7025 |

**TABLE 2.** Optimized energy (O.E.), HOMO- LUMO Energy levels, Chemical Potential (μ), Chemical hardness (η), Electronegativity (χ) and Electrophilicity (ω) values of the studied compounds obtained at B3LYP/LANL2DZ level.

| Cluster | O.E. (Hartree) | $E_{LUMO}$ (eV) | $E_{HOMO}$ (eV) | μ (eV) | η (eV) | χ (eV) | ω (eV) |
|---|---|---|---|---|---|---|---|
| Linear (AsSiTeB) | -42.7344 | -4.2809 | -6.6565 | -5.4687 | 1.1878 | 5.4687 | 12.5891 |
| Bent (SiAsBTe) | -42.8643 | -3.4085 | -6.1868 | -4.7977 | 1.3892 | 4.7977 | 8.2847 |

**Projected Density of States (PDOS) of AsSiTeB/SiAsBTe clusters-**

As the cluster has a high band gap energy then that type of cluster is chemically inert because it requires more energy for the transition of the electron from the HOMO level to the LUMO level for making it to the conductor region. The electrical conduction is low for the clusters with high band gap. Table 1 and Table 2 shows the HOMO-LUMO gap of the AsSiTeB/SiAsBTe structures.

On Analyzing the gap we find that bent structure has higher gap energy as compared with the linear structure which shows the electronic conduction is lower in bent form.

The density-of-states (DOS) spectrum provides the information about the presence of the charges in the cluster [16].

From the Projected density of states PDOS spectrum it is clearly shown that in the linear structure the occupied orbitals is having the even distribution of the charges and in the virtual or unoccupied orbitals there is a higher number of peaks which indicates that the charges easily move to the virtual orbitals which favors the electronic transition or conduction.

From the PDOS spectrum of the bent structure on compairing with linear structure the density of the charges is less in the homo orbital than in the virtual orbital which gives rise to high gap value in this structure.

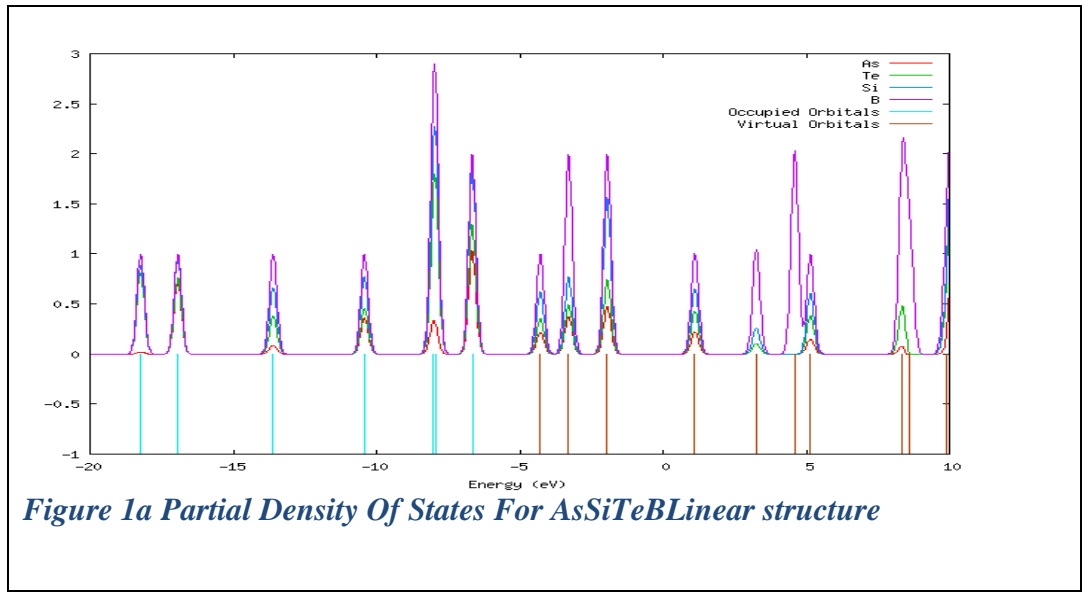

*Figure 1a Partial Density Of States For AsSiTeBLinear structure*

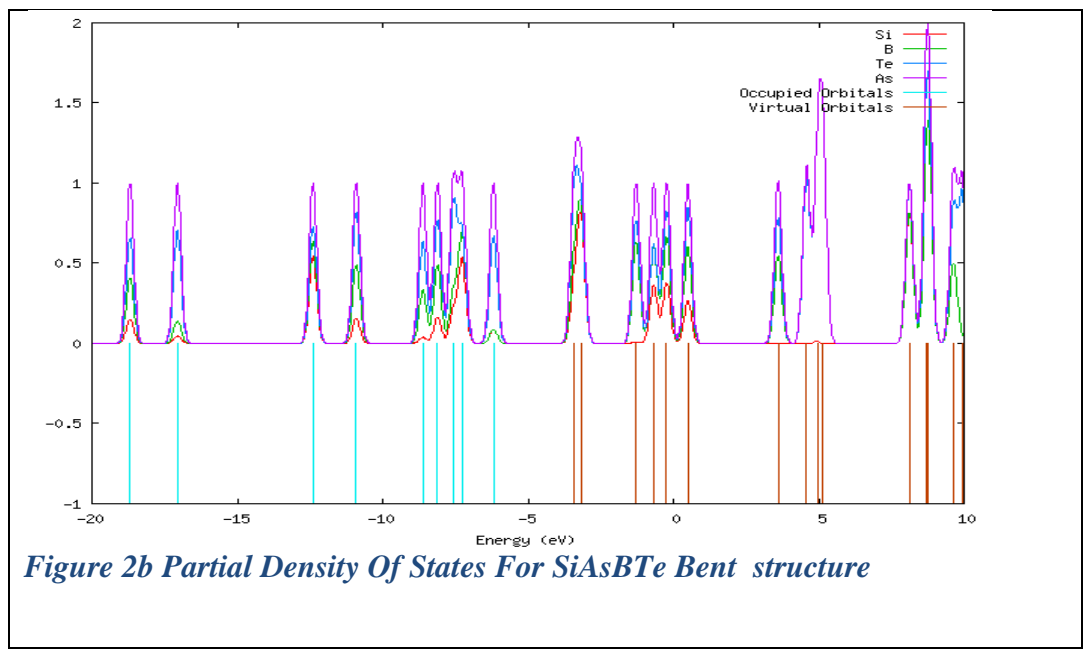

*Figure 2b Partial Density Of States For SiAsBTe Bent structure*

**Vibrational Properties: Infrared And Raman studies on AsSiTeB/SiAsBTe clusters**

The optimized structures for the linear nano-cluster falls into the region of 297.324 micro meter to the 11.169 micro meter at the observed energy of 0.006 eV to the 0.075 eV. This region of the spectrum for the linearly optimized nano cluster signifies the mid infra-red to the far infra-red region. The various observed peaks in the energy spectrum for the infra red to the raman spectrum is also provided in the table 3 below.

The optimized structures for the bent nano-cluster falls into the region of 534.41 micro meter to the 11.92 micro meter at the observed energy of 0.008 eV to the 0.115 eV. This region of the spectrum for the bent form of optimized nano cluster signifies the mid infra-red to the far infra-red region. The various observed peaks in the energy spectrum for the infra red to the raman spectrum is also provided in the table 3 below.

**Table 3- Vibrational Properties**

| Nanoclusters | Properties | Values | Energies (eV) |
|---|---|---|---|
| Bent (SiAsBTe) | **Frequency (Cm$^{-1}$)** | **71.3477, 234.2883, 247.3124, 284.9406, 430.4373, 933.8417** | **0.0088459916, 0.0290480609, 0.0306628442, 0.0353281487, 0.053367449, 0.11578167** |
| | **IR Int.** | **1.4752, 3.581, 1.586, 0.3487, 25.1556, 56.9942** | |
| | **Raman Activity** | **6.3965, 6.4098, 21.387, 9.6138, 47.6248, 18.0927** | |
| Linear (AsSiTeB) | **Frequency (Cm$^{-1}$)** | **-80.0427, -80.427, 55.1146, 55.1146, 152.1337, 512.7741, 583.423** | **-0.00992403471,-0.00992403471, 0.0068333427, 0.0068333427, 0.0188621838, 0.063575916, 0.072335267** |
| | **IR Int.** | **4.057, 4.057, 0.0881, 0.0881, 0.3025, 73.309, 1.8808** | |
| | **Raman Activity** | **57.0067, 57.0067,1.9076, 1.9076, 36.1391, 71.9218, 7.0594** | |

**Conclusion:**

The optimized geometries of the III-V Group semiconductor element (AsSiTeB/SiAsBTe ) nano-cluster have been studied for their energy band gap in the gas phase. The two structures are of linear shape and the other in the bent shape optimized. In which bent shape proves to be more stable than the linear shape. The energy band gap for the linear shape is quite low which can play the role for potential application in the form of light emitting diodes or can have a major role in the application of the solar cells. Therefore this search on the electronic scale and geometrical effect on the cluster is very much useful for the experimentalist to form the new semiconducting alloys. The results of the vibrational analysis suggest us that the present system is very much active in the region of far infra red to the mid infra red region.

**Acknowledgment**

Mr. AnkitKargeti acknowledges BML Munjal University for Ph.D. research fellowship. Dr. Tabish Rasheed acknowledges SERB-DST, Govt. of India for research grant.